\documentclass[journal,final]{IEEEtran}
\IEEEoverridecommandlockouts
\usepackage{cite}
\usepackage{amsmath,amssymb,amsfonts}
\usepackage{bm}
\usepackage{algorithmic}
\usepackage{graphicx}
\usepackage{textcomp}
\usepackage{xcolor,soul}
\usepackage{balance}
\usepackage{comment}

\usepackage[acronym,shortcuts]{glossaries}
\newacronym{af}{AF}{amplify-and-forward}
\newacronym{awgn}{AWGN}{additive white Gaussian noise}
\newacronym{ris}{RIS}{reconfigurable intelligent surface}
\newacronym{mimo}{MIMO}{multiple-input multiple-output}
\newacronym{miso}{MISO}{multiple-input single-output}
\newacronym{ofdm}{OFDM}{orthogonal frequency division multiplexing}
\newacronym{ue}{UE}{user equipment}
\newacronym{sta}{STA}{station}
\newacronym{ap}{AP}{access point}
\newacronym{parafac}{PARAFAC}{parallel factor}
\newacronym{cearc}{CEARC}{channel estimation with adaptive RIS configuration}
\newacronym{mbce}{MBCE}{MUSIC-based channel estimation}
\newacronym{mrc}{MRC}{maximum ratio combining}
\newacronym{mse}{MSE}{mean squared error}
\newacronym[plural=AoAs,firstplural=angles of arrival (AoAs)]{aoa}{AoA}{angle-of-arrival}
\newacronym[plural=AoDs,firstplural=angles of departure (AoDs)]{aod}{AoD}{angle-of-departure}
\newacronym{lmmse}{LMMSE}{linear minimum mean square error}
\newacronym{ls}{LS}{least-square}
\newacronym{zf}{ZF}{zero-forcing}
\newacronym{bs}{BS}{base station}
\newacronym{sinr}{SINR}{signal-plus-interference-to-noise ratio}
\newacronym{snr}{SNR}{signal-to-noise ratio}
\newacronym{ula}{ULA}{uniform linear array}
\newacronym{ce}{CE}{channel estimation}
\newacronym{dft}{DFT}{discrete Fourier transform}
\newacronym{music}{MUSIC}{multiple signal classification}
 \newacronym{omp}{OMP}{orthogonal matching pursuit}
 \newacronym{mmwave}{mmWave}{millimeter wave}
\newacronym{ml}{ML}{machine learning}
\newacronym{los}{LoS}{line-of-sight}
\newacronym{drl}{DRL}{deep reinforcement learning}
\newacronym{d2d}{D2D}{device to device}
\newacronym{nn}{NN}{neural network}
\newacronym{fcnn}{FCNN}{Fully Connected Neural Network}
\newacronym{dt}{DT}{digital twin}
\newacronym{ga}{GA}{genetic algorithm}
\newacronym{rmse}{RMSE}{root mean square error}
\newacronym{csi}{CSI}{channel state information}
\newacronym{pao}{PAO}{prediction-assisted optimization}
\newacronym{sl}{SL}{SINR-based localization}
\newacronym{dtao}{DT-AO}{Digital Twin-Assisted Optimization}
\newacronym{gbo}{GBO}{gradient-based optimization}
\newacronym{upa}{UPA}{uniform planar array}
\newacronym{mmse}{MMSE}{minimum-mean squared error}
\newacronym{rssi}{RSSI}{received signal strength indicator}
\usepackage{tikz}
\usepackage{pgfplots}
\pgfplotsset{compat=newest} 
\pgfplotsset{plot coordinates/math parser=false}
\usetikzlibrary{external}
\usetikzlibrary{plotmarks,patterns,patterns.meta,decorations.pathreplacing,backgrounds,calc,arrows,arrows.meta,spy,matrix}
\usepgfplotslibrary{patchplots,groupplots}
\usepackage{tikzscale}

\usepackage{tikz}
\usepackage{pgfplots}
\usepackage{tikzscale}
\usepackage{tikz-qtree}

\pgfplotsset{compat=newest}
\pgfplotsset{plot coordinates/math parser=false}
\pgfplotsset{every axis/.append style={
                    label style={font=\scriptsize},
                    tick label style={font=\scriptsize},
                    legend style={font=\scriptsize}
                    }}
\usetikzlibrary{plotmarks,shapes,patterns,decorations.pathreplacing,backgrounds,calc,arrows,arrows.meta,spy,matrix,shadows,trees,positioning,fit}
\usepgfplotslibrary{patchplots,groupplots}

\tikzstyle{startstop} = [rectangle, rounded corners, minimum width=2cm, minimum height=0.5cm,text centered, draw=black]
\tikzstyle{io} = [trapezium, trapezium left angle=70, trapezium right angle=110, minimum width=3cm, minimum height=1cm, text centered, draw=black]
\tikzstyle{process} = [rectangle, minimum width=2cm, minimum height=0.5cm, text centered, draw=black, alignb=center]
\tikzstyle{decision} = [ellipse, minimum width=2cm, minimum height=1cm, text centered, draw=black]
\tikzstyle{arrow} = [thick,<->,>=stealth]
\tikzstyle{line} = [thick,>=stealth]
\tikzstyle{darrow} = [thick,<->,>=stealth,dashed]
\tikzstyle{sarrow} = [thick,->,>=stealth]
\tikzstyle{larrow} = [line width=0.1mm,dashdotted,->,>=stealth]

\pgfkeys{/pgf/number format/.cd,1000 sep={\,}} 

\makeatletter
\def\grd@save@target#1{%
  \def\grd@target{#1}}
\def\grd@save@start#1{%
  \def\grd@start{#1}}
\tikzset{
  grid with coordinates/.style={
    to path={%
      \pgfextra{%
        \edef\grd@@target{(\tikztotarget)}%
        \tikz@scan@one@point\grd@save@target\grd@@target\relax
        \edef\grd@@start{(\tikztostart)}%
        \tikz@scan@one@point\grd@save@start\grd@@start\relax
        \draw[minor help lines] (\tikztostart) grid (\tikztotarget);
        \draw[major help lines] (\tikztostart) grid (\tikztotarget);
        \grd@start
        \pgfmathsetmacro{\grd@xa}{\the\pgf@x/1cm}
        \pgfmathsetmacro{\grd@ya}{\the\pgf@y/1cm}
        \grd@target
        \pgfmathsetmacro{\grd@xb}{\the\pgf@x/1cm}
        \pgfmathsetmacro{\grd@yb}{\the\pgf@y/1cm}
        \pgfmathsetmacro{\grd@xc}{\grd@xa + \pgfkeysvalueof{/tikz/grid with coordinates/major step x}}
        \pgfmathsetmacro{\grd@yc}{\grd@ya + \pgfkeysvalueof{/tikz/grid with coordinates/major step y}}
        \foreach \x in {\grd@xa,\grd@xc,...,\grd@xb}
        \node[anchor=north] at (\x,\grd@ya) {\pgfmathprintnumber{\x}};
        \foreach \y in {\grd@ya,\grd@yc,...,\grd@yb}
        \node[anchor=east] at (\grd@xa,\y) {\pgfmathprintnumber{\y}};
      }
    }
  },
  minor help lines/.style={
    help lines,
    gray,
    line cap =round,
    xstep=\pgfkeysvalueof{/tikz/grid with coordinates/minor step x},
    ystep=\pgfkeysvalueof{/tikz/grid with coordinates/minor step y}
  },
  major help lines/.style={
    help lines,
    line cap =round,
    line width=\pgfkeysvalueof{/tikz/grid with coordinates/major line width},
    xstep=\pgfkeysvalueof{/tikz/grid with coordinates/major step x},
    ystep=\pgfkeysvalueof{/tikz/grid with coordinates/major step y}
  },
  grid with coordinates/.cd,
  minor step x/.initial=.5,
  minor step y/.initial=.2,
  major step x/.initial=1,
  major step y/.initial=1,
  major line width/.initial=1pt,
}
\makeatother

\newlength\fheight
\newlength\fwidth

\DeclareMathOperator*{\argmax}{arg\,max}

\usepackage{hyperref}

\usepackage{flushend}
\begin{document}

\title{Digital Twin–Based Beamforming for Interference Mitigation in AF Relay MIMO Systems}

\author{Alexander Bonora, Anna V. Guglielmi,  Davide Scazzoli,
Marco Giordani,~\IEEEmembership{Senior Member, IEEE},\\
Maurizio Magarini,
Vineeth Teeda,
Stefano Tomasin,~\IEEEmembership{Senior Member, IEEE}
\thanks{A. Bonora, A. V. Guglielmi, M. Giordani, and S. Tomasin are with the Department of Information Engineering, University of Padova. Padova, Italy. (Corresponding author: A. Bonora, E-mail: alexander.bonora@phd.unipd.it).\\
D. Scazzoli, M. Magarini, and V. Teeda are with the Department of Electronics, Information and Bioengineering, Politecnico di Milano, Milan, Italy.\\
This work was partially supported by the European Commission through the European Union’s Horizon Europe Research and Innovation Programme under the Marie Skłodowska-Curie-SE, Grant Agreement No. 101129618.\\
This work was partially supported by the European Union
under the Italian National Recovery and Resilience Plan (NRRP) Mission 4,
Component 2, Investment 1.3, CUP C93C22005250001, partnership on
“Telecommunications of the Future” (PE00000001 - program “RESTART”)}
}

\IEEEoverridecommandlockouts
\newcommand\copyrightnotice{%
\begin{tikzpicture}[remember picture,overlay]
\node[anchor=south,yshift=5pt] at (current page.south) {\fbox{\parbox{\dimexpr\textwidth-\fboxsep-\fboxrule\relax}{
\footnotesize \textcopyright 2026 IEEE. Personal use of this material is permitted. Permission from IEEE must be obtained for all other uses, in any current or future media, including reprinting/republishing this material for advertising or promotional purposes, creating new collective works, for resale or redistribution to servers or lists, or reuse of any copyrighted component of this work in other works.
}}};
\end{tikzpicture}
}

\maketitle
\copyrightnotice

\begin{abstract}
Beamforming in \ac{mimo} systems should take interference mitigation into account. However, for beam form design, accurate \ac{csi} is needed, which is often difficult to obtain due to channel variability, feedback overhead, or hardware constraints. For example, \ac{af} relays passively forward signals without measurement, precluding full \ac{csi} acquisition to and from the relay. 
To address these issues, this paper introduces a novel \ac{pao} framework for beam form design in \ac{af} relay-assisted multiuser \ac{mimo} systems. The proposed solution in the \ac{af} relay aims at maximizing the \ac{sinr}. Unlike other methods, \ac{pao} relies solely on received power measurements, making it suitable for scenarios where \ac{csi} is unreliable or unavailable. 
\ac{pao} consists of two stages: a supervised-learning-based \ac{nn} that predicts the positions of transmitters using signal observations, and an optimization algorithm, guided by a \ac{dt}, that iteratively refines the beam direction of the relay in a simulated radio environment. As a key contribution, we validate the proposed framework using realistic measurements collected on a custom-built experimental \ac{mmwave} platform, which enables training of the \ac{nn} model under practical wireless conditions. The estimated information is then used to update the digital twin with knowledge of the surrounding environment, enabling online optimization. Numerical results show the trade-off between localization accuracy and beamforming performance, and confirm that \ac{pao} maintains robustness even in the presence of localization errors while reducing the need for real-world measurements.

\end{abstract}

\glsresetall

\begin{IEEEkeywords}
Amplify-and-Forward (AF) Relay, Beamforming, Digital Twin, Interference, Neural Network. 
\end{IEEEkeywords}

\begin{tikzpicture}[remember picture,overlay]
\node[anchor=north,yshift=-10pt] at (current page.north) {\parbox{\dimexpr\textwidth-\fboxsep-\fboxrule\relax}{
\centering\footnotesize This paper has been submitted to the IEEE Special Issue on "Digital Twins for Wireless Networks: Enabling Application-Aware and Closed-Loop Optimization".}}; 
\end{tikzpicture}

\section{Introduction}
\label{sec:intro}

Dense deployments, multi-user access, and dynamic channel conditions inevitably introduce non-orthogonality into the network, giving rise to interference that can be both severe and unpredictable. To maintain reliable performance, it becomes essential to develop effective interference management techniques tailored to these dynamic environments.

Most existing beamforming techniques in \ac{mimo} systems have been designed to maximize the beamforming gain toward a target user or direction \cite{alkhateeb2014mimo, barati2015directional,giordani2016initial}. Especially when operating in the \ac{mmwave} band, techniques such as discrete Fourier transform (DFT)-based codebooks, hierarchical codebooks, and beam sweeping protocols have been extensively investigated to enhance signal strength while maintaining affordable complexity. These methods have been integrated into wireless standards, including IEEE 802.11ad and 3GPP New Radio (NR), highlighting their practical relevance \cite{ieee80211ad, 3gpp38202}.

However, a key limitation of these approaches is their interference-unaware nature. Specifically, by focusing uniquely on maximizing the desired signal, they neglect the spatial characteristics of potential interfering transmitters, leading to significant performance degradation especially in dense network deployments where multiple links coexist \cite{li2019interference, zhu2016interference}. Indeed, as the density of 
networks increases (with the proliferation of small cells, relays, and user devices), the ability to mitigate interference becomes critical to sustain high quality of service (QoS). This challenge has motivated the investigation of interference-aware beamforming strategies, such as coordinated beamforming and interference alignment, which explicitly account for multi-user interference in the beam design \cite{costa1983dirty, cadambe2008interference}.
Recently, in \cite{zhang2024decentralized} the authors proposed a decentralized interference-aware codebook learning algorithm for multi-cell \ac{mimo} systems that operates without inter-node communication, enabling adaptation in non-stationary environments. The challenge of interference management has also been investigated through location-aware beamforming models that estimate interference effects based on user positioning~\cite{liu2023model}. In \cite{singh2024mitigation}, a survey of new beamforming strategies tailored for \ac{mimo} and code-division multiple-access (CDMA) systems within 5G and beyond architectures was presented.

While these interference-aware methods are promising, they often rely on the availability of either explicit \ac{csi} or accurate channel models to operate effectively. However, acquiring and maintaining up-to-date \ac{csi} is known to be difficult with fast channel variations, high-dimensional antenna arrays, and hardware constraints \cite{choi2016vehicular, alkhateeb2015compressed} among others. Moreover, the overhead associated with channel estimation and feedback can vanish the benefits of interference-aware optimization, especially in dynamic environments with mobility or blockage\cite{Massive_MIMO_CSI}.

To alleviate the dependency of interference mitigation on accurate \ac{csi}, research has explored new solutions based on \ac{ml}, e.g., \cite{huang2019deep, huang2018deep}. Data-driven \ac{ml} models trained on measurement data or simulations, may infer useful channel-related information without explicitly relying on \ac{csi}. A comprehensive study \cite{TRABELSI2024110159} has highlighted their effectiveness in 5G networks.

Recently, \acp{dt}, i.e., virtual replicas of physical systems, have gained attention as tool for simulating, optimizing, and controlling wireless networks in real time \cite{11217298, shamsoshoara2024survey,Paglierani2025DT,pegurri2025van3twin}.
By creating virtual replicas of real-world assets, a \ac{dt} provides an interactive and dynamic platform for analyzing performance, predicting behavior, and facilitating decision-making processes. In the context of communication networks, it is envisioned that \acp{dt} can improve network efficiency by a more efficient and accurate optimization of the main network parameters \cite{vasisht2023digital}. 
Due to the challenges of establishing and then maintaining directional links in a dynamic environment, beam management emerges as a suitable application for \ac{dt}.
In \cite{AlkhateebICC2023}, the authors proposed a \ac{dt}-assisted online interference-aware beam design framework to generate beam patterns that mitigate interference using only power measurements, without requiring explicit channel knowledge. The paradigm of leveraging limited measurements and data-driven optimization motivates also our research work.

Modern networks also include relays to extend coverage and improve reliability, especially in dense multi-user scenarios where interference becomes critical \cite{s20102753}. This issue is particularly acute in \ac{af} relay systems, where full \ac{csi} cannot be obtained for channels from and to the relay, as the relay does not measure signals but simply amplifies and forwards them, precluding direct channel estimation. By forwarding signals, relays may also increase a diversity gain and the overall system robustness. Relays may operate either in a decode-and-forward (DF) mode where the relay restores the received signal, or in an \ac{af}, made when the relay retransmit the received signal without decoding. While DF offers higher reliability, \ac{af} is simpler and more power-efficient~\cite{pagin2022end}. 

Motivated by the limitations of existing literature, we study interference-unaware beamforming in a relay-assisted \ac{mimo} system WiFi scenario. Specifically, we propose an innovative framework, called \ac{pao}, that learns the relay beamforming to maximize the \ac{sinr} without requiring explicit \ac{csi}.

\ac{pao} operates in two stages. First, a \gls{nn} is trained in a supervised fashion to predict the positions of the target user and interfering transmitters based on power measurements. We denote this stage as the \ac{sl}.
The estimated positions are then used in a second stage where an optimization algorithm assisted by the \ac{dt} determines the optimal relay beamforming configuration. We denote this step as \ac{dtao}.
The role of the \ac{dt} is crucial since it is used to emulate the radio channel and evaluate candidate beam directions. To find the relay steering direction that maximizes the \ac{sinr} at the \ac{ap}. This closed-loop approach reduces the need for physical \ac{csi} measurements across different relay beam configurations by leveraging predicted channel information from the \ac{dt}.
The \ac{nn} model is trained and tested using both simulation data generated by a \ac{dt}, and experimental data obtained in realistic wireless conditions from a custom-built \ac{mmwave} platform. This platform captures real-world effects that are difficult to reproduce in simulations, such as hardware imperfections, interference, and environmental variability, making our results more accurate and representative.
Our simulations demonstrate that the proposed approach is robust to non-negligible localization errors. Notably, \gls{pao} achieves a near-optimal solution with around seven times less measurements than the benchmarks.

The remainder of this paper is organized as follows. Sec.~\ref{sec:sisMod} introduces the system model for the relay-assisted MIMO uplink in the presence of non-cooperative interferers. The proposed \ac{pao} framework is presented in Sec.~\ref{sec:promFor}, describing its \ac{sl} stage for position estimation and its \ac{dtao} stage for relay beam refinement. Sec.~\ref{sec:design} details the design of the main system components, including the beamforming codebook, \ac{nn} architecture, digital twin functions $f(\cdot)$ and $f'(\cdot)$, and the employed optimization algorithms, i.e. \ac{gbo} and \ac{ga}. Simulation and experimental results are reported in Sec.~\ref{subsec:num_res}, evaluating both localization accuracy and overall \ac{pao} performance. Finally, Sec.~\ref{sec:concs} concludes the paper.

\paragraph*{\textbf{Notation}} 
Throughout the paper, boldface capital letters and boldface lower-case letters such as $\bm{X}$ and $\bm{x}$ denote matrices and vectors, respectively. $(\cdot)^H$ denotes conjugate transpose. $||\cdot||$ and $|\cdot|$ represent the Frobenius norm and the cardinality of a set, respectively. $\mathbb{E}\{\cdot\}$ denotes the expectation operator.

\section{System Model}
\label{sec:sisMod}
\begin{figure}
    \centering
    \includegraphics[width = \columnwidth]{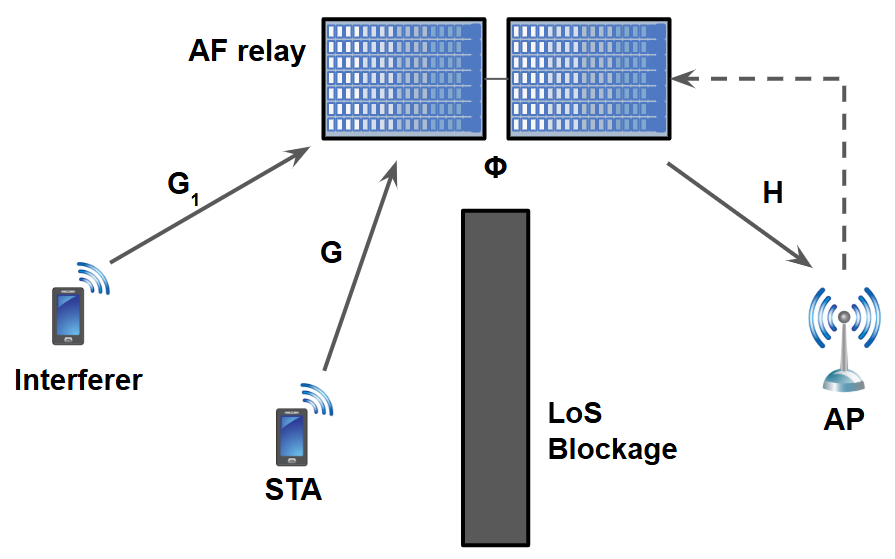}
    \caption{System model. The direct link between the target \ac{sta} to the Wifi \ac{ap} is blocked, so transmission is  assisted by an \ac{af} relay.}
    \label{fig:scenario}
\end{figure}

We consider a narrowband \ac{mimo} uplink transmission system, wherein the transmission from a \ac{sta} with a \ac{upa} of $N_{\rm t}$ antennas to a WiFi \ac{ap}, also equipped with a \ac{upa} of $N_{\rm r}$ antennas, is assisted by an \ac{af} relay as shown in Fig.~\ref{fig:scenario}. In the vicinity of the \ac{ap}, there are \( K > 1 \) non-cooperative devices, where device \( k \in \{1, \ldots, K\} \) has \( N_k \) antennas, operate at the same frequency bands and, therefore, causing interference. 

We adopt a three-dimensional Cartesian coordinate system to describe the network geometry.
Let
\begin{equation}
    \bm{p}_0 = [x_0, y_0, z_0]
\end{equation}
denote the position of the target \ac{sta}, and
\begin{equation}
    \bm{p}_k = [x_k, y_k, z_k], \quad k = 1, \ldots, K,
\end{equation}
denote the positions of the $K$ non-cooperative interfering transmitters respectively.
The positions of the \ac{af} relay and the \ac{ap} are assumed to be fixed and known.
The relative geometry between the \ac{sta}, the interferers, the relay, and the \ac{ap} determines the propagation paths associated with each link.

The \ac{af} relay includes two planar antenna arrays, with $N_\mathrm{i}$ and $N_\mathrm{o}$ antennas, to receive and transmit signals, respectively. The received signals are combined using a relay beamforming vector. The resulting complex scalar signal is then amplified and transmitted after applying a proper precoding beamformer. All channels, beamforming vectors, and signals are expressed in complex baseband equivalent form. The combining and precoding vectors are designed to receive and transmit, respectively, signals from azimuth angles ($\theta_{Az}$) and elevation angles ($\theta_{El}$), Let us define angle vector $\bm{\theta} = [\theta_{Az}, \theta_{El}]$, which contains the geometric propagation paths between the communicating nodes. Hence, to determine the \ac{aoa} and \ac{aod}, as well as the corresponding channel responses, knowledge of the surrounding environment is required. In this work, the environment is assumed to be known and is used to characterize the propagation paths between nodes, enabling a geometry-based channel modeling. The beamforming vector $\bm{w}_N(\theta_{Az}, \theta_{El}) \in \mathbb{C}^{N \times 1}$, where $N = N_x N_y$, is obtained by vectorizing the $N_x \times N_y$ antenna array response:
\begin{equation}
[\bm{w}_N(\theta_{Az}, \theta_{El})]_n = e^{j \frac{2 \pi}{\lambda} (d_x (n_x - 1) \sin \theta_{El} \cos \theta_{Az} + d_y (n_y - 1) \sin \theta_{El} \sin \theta_{Az})}
\end{equation}
for $n = (n_y - 1)N_x + n_x$, with $n_x = 1, \ldots, N_x$ and $n_y = 1, \ldots, N_y$. Then, we denote the resulting phase control matrix of the AF relay (due to the combiner and precoder)~as
\begin{equation}\label{fcris_short}
\bm{\Phi}(\bm{\theta}_i, \bm{\theta}_o) = \mathrm{A}\bm{w}_{N_{\rm o}}(\bm{\theta}_o) \bm{w}^H_{N_{\rm i}}(\bm{\theta}_i),
\end{equation}
where $\mathrm{A}$ is a fixed amplification coefficient. Matrix $\bm{\Phi}$ identifies the {\em relay configuration}. 
We assume that the communication between the target \ac{sta} and the \ac{ap} only happens through the relay, while the \gls{los} link is not available.
We define $\bm{G} \in \mathbb{C}^{N_i \times N_{\rm t}}$, $\bm{G}_k \in \mathbb{C}^{N_i \times N_k}$, and $\bm{H} \in \mathbb{C}^{N_r \times N_o}$ the channels between the \ac{sta} and the relay, the $k$th interfering device and the relay, and the relay and the \ac{ap}, respectively.
The resulting cascaded channels from the target \ac{sta} and the $k$th interferer to the \ac{ap} are
\begin{equation}
\bm{C} = \bm{H} \bm{\Phi}(\bm{\theta}_i, \bm{\theta}_{\rm o}) \bm{G} \quad \text{and} \quad \bm{C}_k = \bm{H} \bm{\Phi}(\bm{\theta}_i, \bm{\theta}_{\rm o}) \bm{G}_k,
\end{equation}
respectively.
To model the received signal and emphasize the dependence on the unknown relay combining vector, we define the effective channel components as
\begin{equation}
h = \bm{w}^H_{N_{\rm r}} \bm{H} \mathrm{A}\bm{w}_{N_{\rm o}}(\bm{\theta}_{\rm o}),  
\end{equation}
\begin{equation}
\bm{g} = \bm{G} \bm{w}_{N_{\rm t}}(\bm{\theta}_{t}), \quad  \bm{g}_{k} = \bm{G}_k \bm{w}_{N_{k}}(\bm{\theta}_{k}),
\end{equation}
where  $\bm{w}_{N_{\rm o}}(\bm{\theta}_{\rm o}) \in \mathbb{C}^{N_{\rm o} \times 1}$, $\bm{w}_{N_{\rm t}}(\bm{\theta}_{\rm t}) \in \mathbb{C}^{N_{\rm t} \times 1}$, and $\bm{w}_{N_{k}}(\bm{\theta}_{k})$ are the combining vector at the \ac{ap} and the precoding vectors at the \ac{sta} and $K$ interference transmitters, respectively.

The relay-to-\ac{ap} link is assumed fixed, with the relay precoder 
$\bm{w}_{N_{\rm o}}(\bm{\theta}_{\rm o})$ pointing directly toward $\bm{p}_{\rm r}$
and the \ac{ap} combiner 
$\bm{w}_{N_{\rm r}}(\bm{\theta}_{\rm r})$ both are pointing directly toward $\bm{p}_{\rm o}$
pre-optimized under no interference on this link, and thus fixed. The \ac{aoa}/\ac{aod} are derived from the relative 3D geometry. 
For a link from a transmitter at $\bm{p}_A$ to a receiver at $\bm{p}_B$, 
the direction vector is 
\begin{equation}
\bm{\Delta} = \bm{p}_B - \bm{p}_A = 
[\Delta x, \Delta y, \Delta z]^\top.
\end{equation}
The corresponding spherical angles are
\begin{equation}
\theta_{\rm El} = \arcsin\Bigg(\frac{\Delta z}{|\bm{\Delta}|}\Bigg), \quad \theta_{\rm Az} = \mathrm{atan2}(\Delta y, \Delta x),
\end{equation}
where $|\bm{\Delta}| = \sqrt{\Delta x^2 + \Delta y^2 + \Delta z^2}$ 
is the Euclidean distance, $\arcsin(\cdot) \in [-\pi/2, \pi/2]$, 
and $\mathrm{atan2}(\Delta y, \Delta x) \in [-\pi, \pi]$ gives the azimuthal angle 
in the $xy$-plane measured from the $x$-axis toward the $y$-axis. Furthermore, the \ac{sta} \ac{upa} is assumed to be oriented toward the relay, 
so the \ac{sta} precoding vector is optimized and fixed. 
However, the combining vector at the relay remains unknown and is to be designed.

Accordingly, as the \ac{sta} transmits a symbol $x \in \mathbb{C}$ to the \ac{ap}, and the other $K$ interfering devices also transmit symbols $x_k \in \mathbb{C}$, $k=1,...,K$, in the same time and frequency slot, the received signal at the \ac{ap} can be expressed as
\begin{equation}\label{recSig}
        y =  h\bm{w}^H_{N_{\rm i}}(\bm{\theta}_{\rm i})\bm{g} x +  \sum_{k=1}^{K} h\bm{w}^H_{N_{\rm i}}(\bm{\theta}_{\rm i})\bm{g}_k x_k +  n \,,
\end{equation}
where $n \sim \mathcal{CN}(0, \sigma^2)$ is the \ac{awgn} with zero-mean and power $\sigma^2$. 

\section{Prediction Assisted Optimization Framework for Relay Configuration}
\label{sec:promFor}

In this section, we present the proposed \ac{pao} framework for relay configuration, which enables efficient beam optimization in interference-limited environments without requiring full \ac{csi} acquisition. As shown in Fig.~\ref{fig:model}, the framework integrates four main functional blocks: the Measurement Module, the Switching Control Unit, the \ac{sl} Module, and the \ac{dtao} component. These components jointly support localization-aware beam optimization and low-overhead communication.
The overall operation alternates between a measurement phase and a communication phase, orchestrated by the Switching Control Unit in Fig.~\ref{fig:model}. During measurement, signal observations are collected and processed for environment inference, while during communication, the optimized relay configuration is applied for data transmission.

\paragraph{Measurement Module}
\label{subsec:measurement}

As illustrated in Fig.~\ref{fig:model}, the Measurement Module is responsible for acquiring received signal power and interference-plus-noise power samples under different relay beam configurations. In measurement mode, a predefined codebook of beam directions is sequentially applied at the relay based on the signaling received from the \ac{ap} through a control link, and the corresponding \ac{sinr} values are computed directly from the measured powers without explicit channel estimation.
This lightweight measurement process avoids the overhead of full \ac{csi} acquisition while still capturing the spatial and interference characteristics of the environment. The resulting \ac{sinr} observations are forwarded to the \ac{sl} Module for further processing, as indicated in Fig.~\ref{fig:model}.

\paragraph{Switching Control Unit}
\label{subsec:switching}

The Switching Control Unit serves as the central coordinator of the framework, as depicted in Fig.~\ref{fig:model}. It dynamically switches the relay operation between measurement mode and communication mode based on control commands transmitted over a dedicated control link from the \ac{ap}.
In measurement mode, the Switching Control Unit activates the beam sweeping process and triggers the Measurement Module to collect \ac{sinr} samples. Once sufficient measurements are obtained and an optimized configuration is identified, it transitions the relay to communication mode by applying the selected beam configuration to the RF chain. At the same time, it disables the measurement process and the \ac{dt}-based optimization to minimize computational and signaling overhead, as shown in Fig.~\ref{fig:model}.

\paragraph{\ac{sl} Module}
\label{subsec:sl}

The \ac{sl} Module processes the measured \ac{sinr} values provided by the Measurement Module in Fig.~\ref{fig:model}. It employs an \ac{nn} to infer the spatial configuration of the environment, including the estimated positions of the target \ac{sta} and interfering transmitters.
By learning the mapping between beam-dependent \ac{sinr} patterns and spatial layouts, the \ac{sl} Module enables environment-aware decision making without explicit channel reconstruction. These predicted positions serve as inputs to the \ac{dtao} component when further beam refinement is required, as illustrated in Fig.~\ref{fig:model}.

\paragraph{\ac{dtao}}
\label{subsec:dtao}

The \ac{dtao} component leverages a \ac{dt} as a virtual representation of the physical environment, as shown in Fig.~\ref{fig:model}. Using the position estimates provided by the \ac{sl} Module, the \ac{dt} emulates propagation conditions and interference effects to iteratively evaluate candidate relay beam configurations.
Through simulated \ac{sinr} feedback, the \ac{dtao} component searches for a beam direction that maximizes link quality under the inferred environment conditions. Once a suitable configuration is found, the optimized beam parameters are sent to the Switching Control Unit, which applies them during communication mode as depicted in Fig.~\ref{fig:model}.

 In this way, the \ac{dt} is only engaged when optimization or reconfiguration is required, while regular communication relies solely on the selected relay configuration, resulting in a practical and resource-efficient architecture that tightly integrates learning, optimization, and hardware control.

\subsection{Problem Formulation}
\label{sub:problem-formulation}
We defined the received signal at the \ac{ap} in~\eqref{recSig}. Based on this received signal model, and for notational simplicity, in the remainder of this section we denote the relay combining vector as $\bm{w}_{N_{\rm i}}(\bm{\theta}_i) = \bm{w}(\bm{\theta})$. Under this notation, the resulting \ac{sinr} at the \ac{ap} can be expressed as
\begin{equation}\label{sinr}
    \gamma =
    \frac{\left| h\, \bm{w}^H(\bm{\theta})\, \bm{g} \right|^2 P_x}
    {\sum_{k=1}^{K} \left| h\, \bm{w}^H(\bm{\theta})\, \bm{g}_k \right|^2 P_{x_k} + \sigma^2}.
\end{equation}
where $P_x$ denotes the transmit power of the target \ac{sta}, and $P_{x_k}$ is the transmit power of the $k$th interfering signal.

The goal is configuring the relay to maximize the rate. From the received signal~\eqref{sinr}, the achievable channel rate between the target \ac{sta} and the \ac{ap} is
\begin{equation} \label{rate}
    R = \log_2 ( 1 + \gamma(\bm{\theta})),
\end{equation}
Note that, since the relay precoding vector is fixed, optimization is performed solely over the relay combining vector. Since the combining vector is defined as a steering vector, its optimization comes down to finding the optimal azimuth and elevation angles, i.e., $\bm{\theta} {=} [\theta_{Az}, \theta_{El}]$, that maximize the \ac{sinr}. Therefore, we aim at solving the following problem:
\begin{equation}
\label{optPrSINRfixed}
\hat{\bm{\theta}} = \argmax_{\bm{\theta}}\gamma(\bm{\theta}).
\end{equation}

In principle, this optimization can be easily performed with the assumption that full \ac{csi} is available. However, in practice, obtaining such information is challenging due to the large training overhead required for estimating multiple channel components, particularly in the presence of a relay\cite{8399065}. The \ac{af} relay cannot measure the channel to the \ac{sta} or to the \ac{ap}, since it does not include a signal processing unit; thus such information can only be obtained by the \ac{ap} through lengthy estimates at the cascade channel with several relay configurations.
However, note that~\eqref{sinr} depends solely on the power levels of the received and interference signals, and we will aim at maximizing $\gamma$ using only power measurements. In particular, we measure the received \ac{sta} power $| h\, \bm{w}(\bm{\theta})\, \bm{g}|^2 P_x$, and the combined interference-plus-noise power $\sum_{k=1}^{K} \left| h\, \bm{w}(\bm{\theta})\, \bm{g}_k \right|^2 P_{x_k} + \sigma$. Hence, we aim at optimizing the relay beam direction $\bm{\theta}$, using only these measured powers.

\begin{figure*}[t]
    \centering
    \includegraphics[width=\textwidth]{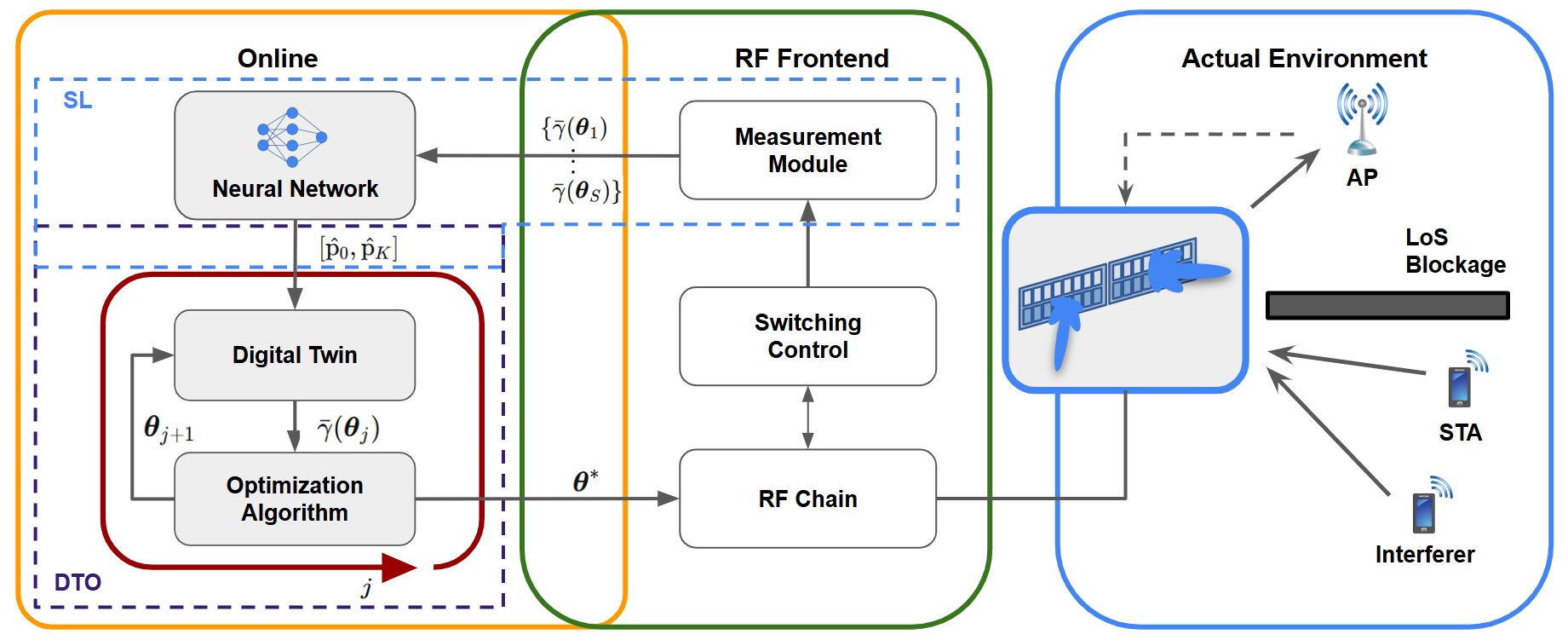}
    \caption{Block diagram of the proposed \ac{pao} framework for relay beam configuration. The architecture integrates a Measurement Module, Switching Control unit, \ac{sl}-based localization block, and \ac{dt}-assisted optimization engine, enabling dynamic switching between measurement and communication modes to achieve efficient beam optimization without full \ac{csi} acquisition.}
    \label{fig:model}
\end{figure*}

To this end, we propose a two-stage framework called \ac{pao}. In the first stage (\ac{sl}, described in Sec.~\ref{sl}), an \ac{nn} is trained to estimate the positions of the \ac{sta} and the interfering devices based on \ac{sinr} measurements.
The second step (\ac{dtao}, described in Sec.~\ref{vo}) optimizates the relay combining beam direction. This is accomplished through an iterative loop between a \ac{dt} and an optimization algorithm, where the combining beam direction $\bm{\theta}$ is progressively refined using environmental insights provided by the \ac{dt}. The process operates in a closed loop, continuously updating $\bm{\theta}$ to maximize $\gamma$.
Figure ~\ref{fig:model} shows the overall scheme of the \ac{pao} framework. The use of \ac{dt} enables to significantly reduce the number of \ac{csi} estimates (in the first stage) and still being able to infer the cascade channel for only relay configuration. This then enables computing the optimal configuration by explaining the digitally emulated environment.

\subsection{\texorpdfstring{\acrfull{sl}}{sl}}\label{sl}

The \ac{sl} constitutes the first step of the proposed \ac{pao} framework, and aims at estimating the \ac{mmse} positions of the target \ac{sta} and interfering transmitters based on observed \ac{sinr} measurements. We first define as the concatenation vector of the 3D coordinates of all $K+1$ transmitters.
\begin{equation}
    \bm{p} = [\bm{p}_0, \bm{p}_1, \dots, \bm{p}_K] \in \mathbb{R}^{3(K+1)}
\end{equation}

To collect inputs for the \ac{sl} step, a set of $S$ different combining beam directions at the relay are explored. Each direction is characterized by a steering vector $\boldsymbol{\theta}_i$, chosen from the predefined codebook of cardinality $C$, $\mathcal{C} = \{\boldsymbol{\theta}_1, \dots, \boldsymbol{\theta}_C\}.$ For each selected beam direction $\boldsymbol{\theta}_i$, the corresponding \ac{sinr} value $\bar{\gamma}(\boldsymbol{\theta}_i)$ is measured. These measurements are collected in the set $\bm{\bar{\Gamma}} = [\bar{\gamma}(\boldsymbol{\theta}_1), \dots, \bar{\gamma}(\boldsymbol{\theta}_S)]$. The \ac{sl} step takes $\bar{\Gamma}$ as input, and outputs an estimate $\hat{\bm{p}}$ of the true device (\ac{sta} and interferers) positions. The localization problem can be formulated as
\begin{equation}\label{msedef}
    \hat{\bm{p}} = \arg \min_{\bm{p}} \mathbb{E} \left[ \|\bm{p} - \bar{\bm{p}}\|^2 \,\big|\, \bm{\bar{\Gamma}} \right],
\end{equation}
where $\bar{\bm{p}}$ denotes the true positions of the transmitting devices, and $\hat{\bm{p}}$ is the estimated output of the \ac{sl} step.

Since~\eqref{sinr} is affected by both the signal strength from the target \ac{sta} and that from the interference transmitters, it is used to determine the spatial coordinates of the transmitting devices. Specifically, a mapping from the  measurements collected in $\bar{\Gamma}$ to the spatial coordinates $\bar{\bm{p}}$ of the transmitters' positions need to be learned. Since this mapping is non linear, we resort to an \ac{nn} for the prediction. The loss function used for training is the \ac{mse} in~\eqref{msedef}.

For the training of the \ac{nn} in the \ac{sl} step, we adopt a two-phase approach. We first pre-train the model using a simulated dataset
\begin{equation}\label{simdata}
    {\mathcal D}_S = \{ (\bm{\bar{\Gamma}}, \bar{\bm{p}})_n \}_{n=1}^{|{\mathcal D}_S|}\,.
\end{equation}
generated by a \ac{dt}.
The details of the simulation setup will be provided in Section~\ref{subsec:setup}.
In the second phase, the model is {\em fine-tuned} using experimental data to adapt to real-world conditions. Specifically, for the fine-tuning we employ the dataset
\begin{equation}\label{expdata}
    {\mathcal D}_T = \{ (\bm{\tilde{\Gamma}}, \bar{\bm{p}}) \}_{n=1}^{|{\mathcal D}_T|}.
\end{equation}
with experimental data obtained from a real laboratory setup (as will be described in Sec.~\ref{sec:sub_experimental_setup}).
This fine-tuning ensures that the pre-trained \ac{nn} is effectively calibrated to the environment where it is deployed.
Details about the structure of the NN architecture will be provided in Sec.~\ref{subsec:design_nn}.

\subsection{\texorpdfstring{\acrfull{dtao}}{DTAO}}\label{vo}
The \ac{dtao} is the second step of the proposed \ac{pao} framework. It consists of an iterative  procedure between an optimization algorithm and the \ac{dt}. The objective of \ac{dtao} is to optimize the steering direction $\bm{\theta}$ of the relay combining vector to improve the \ac{sinr} at the target \ac{sta}, given the positions of the target \ac{sta} and the interference from the transmitters that is predicted by the \ac{sl} step.
At this stage, the predicted positions vector $\hat{\bm{p}}$, as well as a candidate steering direction $\bm{\theta}$, are the inputs of the \ac{dt} that act as a virtual representation of the actual environment. In particular, given  $\hat{\bm{p}}$, the \ac{dt} evaluates the impact of a candidate steering direction $\bm{\theta}_j$ provided by an optimization algorithm, which will be further described in Sec.~\ref{subsec:design_oa}, at iteration $j \geq 1$,
by first computing the received signal power 
\begin{equation}\label{ue_signal}
\hat{P}_S = f(\hat{\bm{p}}_0, \bm{\theta}_j)
\end{equation}
and the interference-plus-noise power
\begin{equation}\label{interf_signal}
\hat{P}_{I+N} = f'(\hat{\bm{p}}_k, \bm{\theta}_j)\,,
\end{equation}
for $k=1, \ldots, K$, where functions $f(\cdot)$ and $f'(\cdot)$. These two functions will be described in Sec.~\ref{sec:design_digital_twin}. It is possible to then evaluate the resulting \ac{sinr}
\begin{equation}\label{sinr_position}
\hat{\gamma}(\bm{\theta}_j) = \frac{\hat{P}_S}{\hat{P}_{I+N}}.
\end{equation}
To compute \eqref{ue_signal} and \eqref{interf_signal} the \ac{dt} must be able to reproduce the effects of the physical properties of the environment, ensuring a realistic simulation of signal propagation.  
The value of $\hat{\gamma}(\bm{\theta}_j)$ is then used by the optimization algorithm to guide the selection of the next candidate steering direction $\bm{\theta}_{j+1}$.
We obtain a closed-loop procedure, where the optimization algorithm iteratively queries the \ac{dt} with updated steering directions $\bm{\theta}$, and uses the returned \ac{sinr} values $\hat{\gamma}$ to refine its search. The process continues until convergence (i.e., when no further improvement of the \ac{sinr} is observed) or when a maximum number of iterations $J$ is reached.

\paragraph*{Remark} The optimization process is driven entirely by simulated feedback from the \ac{dt} rather than using costly real-world measurements. By relying on the \ac{dt} as a surrogate model for signal propagation and interference, the optimization algorithm can rapidly explore the steering direction space, and converge toward an \ac{sinr}-optimal relay configuration requiring few interactions with the environment.

\section{Design of the PAO Framework Components}
\label{sec:design}

In this section, we describe the design of the specific components of the proposed \ac{pao} framework introduced in Sec.~\ref{sec:promFor}, which leverages position predictions to guide beam configuration decisions at the relay. In the next subsection~\ref{sub:codebook} we present the design of the configuration codebook $\mathcal{C}$ used in the \ac{pao} \ac{sl} step. This is followed by the description of the \ac{nn} architecture employed for position prediction in~\ref{subsec:design_nn}.
Next, in Sec.~\ref{sec:design_digital_twin} we introduce the functions $f(\cdot)$ and $f'(\cdot)$ that predict the received \acp{sta} powers of a candidate beam steering direction $\bm{\theta}$. Finally, in Sec.~\ref{subsec:design_oa} we explain the optimization algorithm that selects the beam direction maximizing the expected \ac{sinr} according to the predicted positions.

\subsection{Codebook Design}
\label{sub:codebook}

To construct codebook $\mathcal{C}$, we quantize the angular space into finite sets of azimuth and elevation angles, consisting of $M$ elevation angles and $N$ azimuth angles, based on the resolution limits imposed by the phase shifters. Specifically, we define the sets of quantized angles as $\Theta = \{\theta_1, \dots, \theta_M\}, \Xi = \{\xi_1, \dots, \xi_N\}$,
where $\theta_{m+1} - \theta_m = \Delta_\theta$ and $\xi_{n+1} - \xi_n = \Delta_\xi$. The codebook $\mathcal{C}$ is then constructed as the Cartesian product of these two angle sets, i.e.,\begin{equation} \mathcal{C} = \left\{ \left( \theta_{m}, \xi_{n} \right) \right\}^{M \times N}. \end{equation} Hence, the total number of codebook entries is $C = M\cdot N$.

\subsection{Design of the NN}
\label{subsec:design_nn}
The \ac{nn} is used in the \gls{pao} \gls{sl} step to determine transmitter's and interferers' spatial positions. This is done by mapping a vector of measured \acp{sinr} to the spatial coordinates of the \ac{sta} and interferers. The architecture of the \ac{nn} includes $ L $ layers, each composed of $ B_\ell $ neurons, $ \ell \in \{1, \dots, L\} $, and is designed to learn the mapping from observed features to the predicted positions. Specifically, the $ B_{\ell-1} \times 1 $ vector $ \bm{F}_\ell $ is the input of layer $ \ell $, where $ B_\ell $ is the input dimension of layer $ \ell $. For the first layer ($ \ell = 1 $), we have
$\bm{F}_1 = \bm{\bar{\Gamma}}$, i.e., the vector of $S$ \ac{sinr} measurements. The output of the $ \ell $th hidden layer is calculated as $\bm{F}_{\ell+1} = g_\ell(\bm{W}_\ell \bm{F}_\ell + \bm{b}_\ell),$ where $ \bm{W}_\ell $ (of size $ B_\ell \times B_{\ell-1} $) and $\bm{b}_\ell $ (of size $ B_{\ell} \times 1 $) are the trainable weights and biases, respectively, and $ g_\ell $ is the activation function for layer $ \ell $. The output of the last layer ($ \ell = L $) returns the vector of the predicted~positions $\hat{\bm{p}} = g_L(\bm{W}_L \bm{F}_L + \bm{b}_L).$
The activation function in the last layer $g_L(\cdot)$ is typically chosen to fit the desired output properties. Since positions are continuous real values, a linear activation function is often used.

To improve generalization, the \ac{nn} is first pre-trained on dataset $ \mathcal{D}_S $, allowing it to learn a broad representation of the mapping from \ac{sinr} to positions. The model is then refined on the dataset $ \mathcal{D}_T $ made of real \ac{sinr} measurements to adapt to domain-specific characteristics. During refinement, some layers are frozen, keeping their weights fixed to retain previously learned representations, while the remaining layers are fine-tuned on $ \mathcal{D}_T $.

Since we aim at minimizing the difference between he actual and predicted positions of the target \ac{sta} and the interfering transmitters from we use the \ac{mse} loss function to train the \ac{nn}, the training aims at minimizing, i.e.,
\begin{equation} \label{mse_formula}
     {\rm MSE}(\bar{\bm{p}}, \hat{\bm{p}}) = \frac{1}{||{\mathcal D}||} \sum_{ \bar{\bm{p}} \in {\mathcal D}}{ ||\hat{\bm{p}} - \bar{\bm{p}}||^2}\,,
\end{equation}
where ${\mathcal D}$ is either ${\mathcal D}_S$ or ${\mathcal D}_T$ according to the training~phase.

\subsection{Design of Functions \texorpdfstring{$f(\cdot)$}{f(.)} and \texorpdfstring{$f'(\cdot)$}{f'(.)}}
\label{sec:design_digital_twin}
The objective of the \ac{dt} in the \gls{pao} \gls{dtao} step is to simulate signal propagation and interactions in a realistic wireless environment. We employ MATLAB as a simulation engine for simplicity. First the digital beamformer is modeled using the phased array toolbox \cite{mathworks_phased_array_toolbox}, multiple of these are then integrated into a site-viewer object which creates a 3D environment modeled after the real office used for the experimental measures. Finally, raytracing is also performed in MATLAB, using the communications toolbox. Using these tool we implemented the functions $f(\cdot)$ and $f'(\cdot)$. These functions take as inputs the estimated positions of the \ac{sta} and interferers (vector $\bm{p}$) and a beam steering direction $\bm{\theta}$. The simulator constructs a complete transceiver system, including the antenna array geometry, scene layout, and channel characteristics. The function $f(\bm{p}, \bm{\theta})$ processes this geometry and steering direction to compute the received signal at the \ac{sta}. In parallel, $f'(\hat{\bm{p}}_k, \bm{\theta}_j)$ calculates the aggregate interference and noise power by summing contributions from interferers at positions $\hat{\bm{p}}_k$ under steering direction $\bm{\theta}_j$.

\paragraph*{Remark} While a \ac{dt} can be built using ray tracing to emulate the physical environment, this approach often is hard to implement. The complexity of tracing electromagnetic waves, combined with partial knowledge of material and simplified environmental models~\cite{lecci2021accuracy}, limits the accuracy and scalability of ray-tracing-based \acp{dt} \cite{ying2025site}. In contrast, data-driven surrogate models, such as \acp{nn}, can be trained directly on measured or simulated data, eliminating the need for detailed modeling of the environment \cite{pr9030476}. This approach simplifies development but may be demanding in computational resources. Regardless of the method used to design the \ac{dt}, the key advantage remains the same: the ability to optimize the precoding vector effectively with significantly fewer real-world interactions, providing a more efficient and cost-effective alternative to traditional approaches based on real measurements. In this work, a ray-tracing-based \ac{dt} is adopted due to environment information availability and its relatively lower computational cost compared to large-scale data-driven training pipelines, making it a practical and reproducible choice for the considered scenario.

\subsection{Design of the Optimization Algorithm}
\label{subsec:design_oa}
To find the optimal beam steering direction $\bm{\theta}$ in the \gls{pao} \ac{dtao} step by~\eqref{optPrSINRfixed}, we consider two optimization algorithms: the \ac{gbo} algorithm and the \ac{ga}. 
\paragraph*{Gradient-Based Optimization (GBO)}  
This method employs a multi-start interior-point algorithm initialized from multiple starting points, denoted by $D_{MS}$, which correspond to different initial beam steering directions $\bm{\theta} \in \{\bm{\theta}_1, \bm{\theta}_2, \dots, \bm{\theta}_{D_{MS}} \}$.
These directions are considered to thoroughly explore the solution space and mitigate the risk of converging to poor local minima. For each initial steering direction $\bm{\theta}_d$, with $d=1, \ldots, D_{MS}$, the constrained nonlinear multivariable function is maximized, resulting in a local maximum $\hat{\bm{\theta}}_d$. Since multiple initial points can lead to the same local maximum, the set of the obtained unique local maximum is denoted by $\mathcal{M} = \{ \hat{\bm{\theta}}^{(1)}, \hat{\bm{\theta}}^{(2)}, \ldots, \hat{\bm{\theta}}^{(K)} \}, \, K \leq D_{MS}$.

Among the $K$ unique local minima in $\mathcal{M}$, the solution of~\eqref{optPrSINRfixed} is obtained as
\begin{equation}\label{maxsol}
    \hat{\bm{\theta}}_{\rm GBO} = \argmax_{\hat{\bm{\theta}} \in \mathcal{M}}\hat{\gamma}(\hat{\bm{\theta}}).
\end{equation}
\paragraph*{Genetic Algorithm (GA)}
The \ac{ga} is an optimization method that is well-suited for highly nonconvex and multi-modal optimization problems. 
It is based on the principle of natural selection, where a population of candidate solutions evolves over successive generations. Specifically, at each generation $g$, the population consists of $G$ individuals $\bm{\theta}_g$, $g=1, \ldots, G$, where $\bm{\theta}_g$ represents a candidate beam steering direction.
The fitness of each individual is evaluated based on the objective function, where a higher \ac{sinr} corresponds to better fitness. Through selection, crossover, and mutation operations, the population gradually evolves towards high-performing solutions. Upon convergence, the individual yielding the best fitness is selected as the final output, that is
\begin{equation}\label{eq:ga_sol}
    \hat{\bm{\theta}}_{\rm \text{GA}} = \argmax_{ \{\hat{\bm{\theta}}_g,\, g=1, \ldots, G\} } \hat{\gamma}(\hat{\bm{\theta}}_g)\,,
\end{equation}
where $G$ is the index of the final generation.
Through these evolutionary steps, the population is iteratively updated to promote the emergence of beam steering directions that effectively solve~\eqref{optPrSINRfixed}.

\section{Numerical Results}
\label{subsec:num_res}
In this section, we evaluate and discuss the performance of the proposed \ac{pao} framework for the optimization of the relay configuration.

\subsection{Experimental Setup}\label{sec:sub_experimental_setup}
Our proposed \ac{pao} framework is trained and validated using both simulated and experimental data.
Specifically, we compute and collect the \ac{rssi} as a reference power measurement, for different beam steering directions, to then compute the \ac{sinr}.

\begin{figure}
    \centering
    \includegraphics[width = \columnwidth]{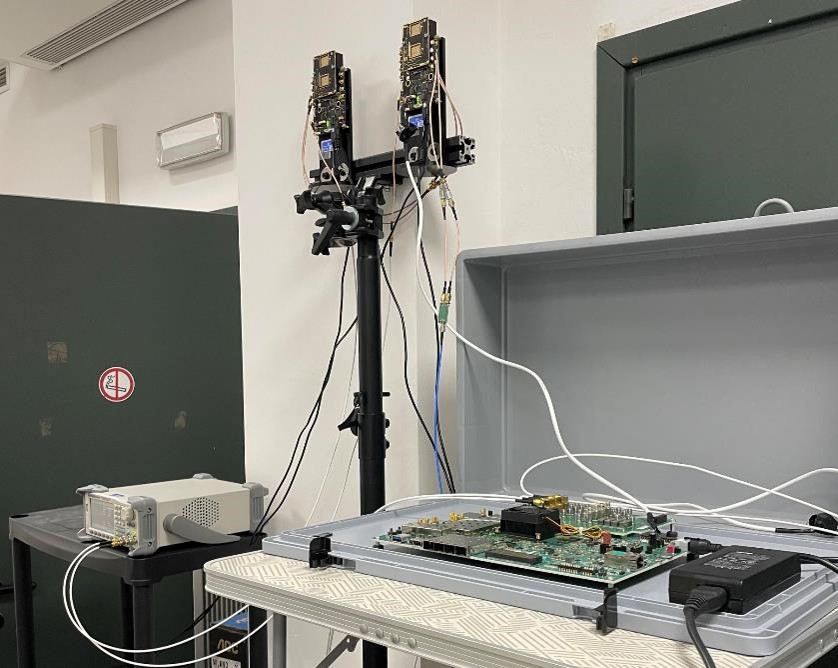}
    \caption{Photograph of the mmWave relay prototype composed of two Sivers EVK06003 beamforming modules, integrating 16-element antennas and RF front-end functionalities for transmission and reception in the 60 GHz band.}
    \label{fig:mmwave_relay}
\end{figure}

The RF \ac{mmwave} front-ends consist of Sivers EVK06003~\cite{Sivers} modules, which integrate beamforming and up/down conversion functionalities using a 16-element URA antenna in a $2\times8$ configuration. The beamformers operate with a predefined codebook that covers a grid of directions with azimuth angles ranging from $-54^\circ$ to $+54^\circ$ in $5.4^\circ$ steps, and elevation angles of $+18^\circ$, $0^\circ$, and $-18^\circ$. The devices support frequency conversion between $59$ and $67$ GHz; for the experiments, we operate in the ISM band at a central frequency $60.48$~GHz, as illustrated in Fig.~\ref{fig:mmwave_relay}.

Baseband signal generation and acquisition are handled by an RFSoC ZCU111 platform~\cite{HW_ZCU1111}, which interfaces with both the \ac{sta} and the AF relay receiver. The interfering transmitter is implemented using a pre-programmed ADALM-Pluto SDR. 

\begin{figure}
\centering 
\includegraphics[width = 0.95\columnwidth]{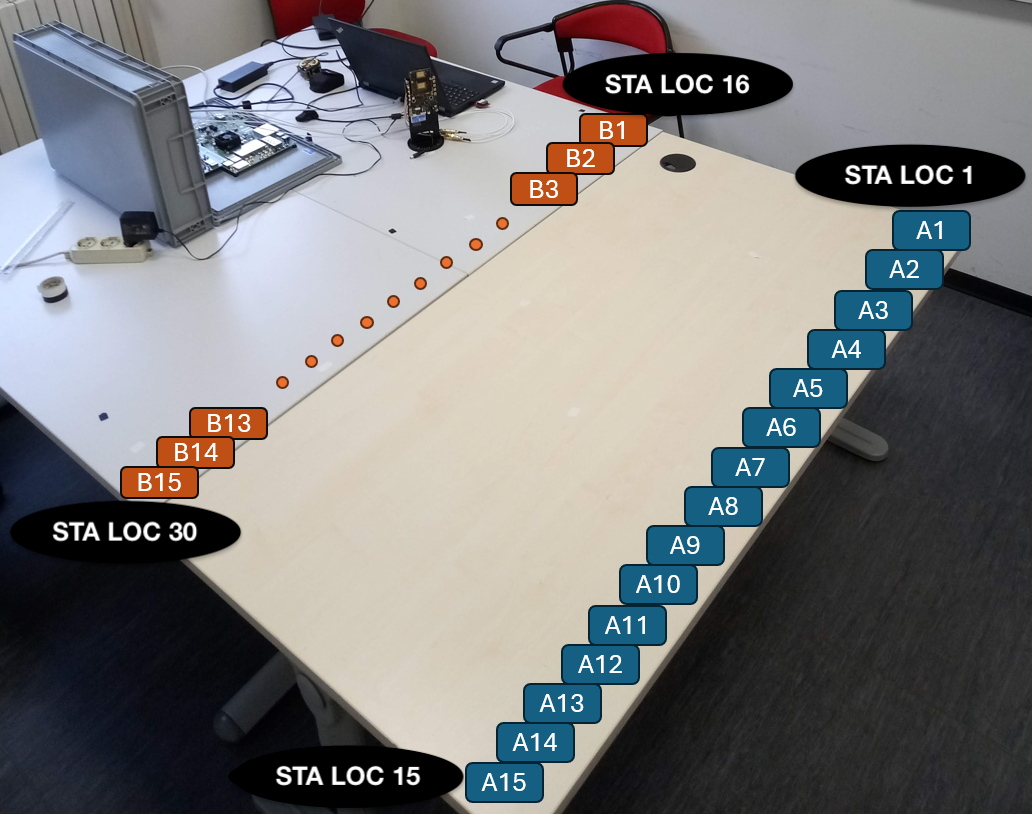}
\caption{The transmitter array consists of 30 positions, organized into two parallel rows along the longer sides of a $1.5~\mathrm{m} \times 1~\mathrm{m}$ rectangle in the $x$-$y$ plane. Each row contains 15 positions, evenly spaced at $0.1~\mathrm{m}$. Positions A and B denote the first and last transmitter locations along each row.}
\label{fig:exper_data_pos}
\end{figure}

Both the \ac{sta} and interfering transmitters are configured to use their optimal beam steering directions, and the alignment with the receiver is achieved using a laser for precise pointing. This alignment simplifies the dataset by reducing beam misalignment variability. As shown in Fig.~\ref{fig:exper_data_pos}, power measurements are collected at the AF relay receiver from 30 distinct positions of the target transmitter, with 20 beam steering direction sweeps per position. Positions are labeled from A1 to A15 and from B1 to B15, and arranged along two parallel lines (A and B) corresponding to the longer sides of a  $1.5  \mathrm{m} \times 1 \mathrm{m}$ rectangle in the $x$-$y$ plane.

\begin{table}[t!]
\centering
\caption{Simulation parameters.}
\label{tab:parameters}
\begin{tabular}{|l|p{2.8cm}|p{2.2cm}|}
\hline
\textbf{Parameter} & \textbf{Description} & \textbf{Value} \\
\hline \hline
$F_0$ & Operating frequency & $60.48$ GHz \\
$\lambda$ & Wavelength & ${c}/{F_0}$ \\
$B$ & Bandwidth & $1.2 \times 10^9$ Hz \\
$N_{\rm t}$, $N_{\rm i}$, $N_{\rm o}, N_{\rm r},N_{\rm k}$ & Antenna array structure & $2 \times 8$ \\
$Tx_{loc}$ & Transmitter xyz position & $[x_{tx}, y_{tx}, 0.9]$ \\
$Rx_{loc}$ & Receiver xyz position & $[9, 2, 1.2]$ \\
$AF_{Rx_{loc}}$ & AF receiver relay xyz position & $[7, 5.5, 2]$ \\
$AF_{Tx_{loc}}$ & AF transmitter relay xyz position & $[7.2, 5.5, 2]$ \\
$I_{loc}$ & Interferer position & $[x_{int}, y_{int}, 0.9]$ \\
$x_{min}$, $x_{max}$ & Min-max range for $x$ & 3.4 - 6.5 m\\
$y_{min}$, $y_{max}$ & Min-max range for $y$ & 1 - 2.9 m\\
$x_{tx}, y_{tx}$ & Transmitter \mbox{coordinates} & $\in [x_{min}, x_{max}] \times [y_{min}, y_{max}]$ \\
$x_{int}, y_{int}$ & Interferer \mbox{coordinates} & $\in [x_{min}, x_{max}] \times [y_{min}, y_{max}]$ \\
$P_{Tx}$ & Transmitter power& $0.1$ W \\
$P_i$ & Interferer power & $0.1$ W \\
$\Theta$ & Azimuth angle & $\in[-54^\circ,54^\circ]$ \\
$\Xi$ & Elevation angle & $\{-18, 0, 18\}$ \\
\hline
\end{tabular}
\end{table}

\begin{figure}
    \centering
    \includegraphics[width = \columnwidth]{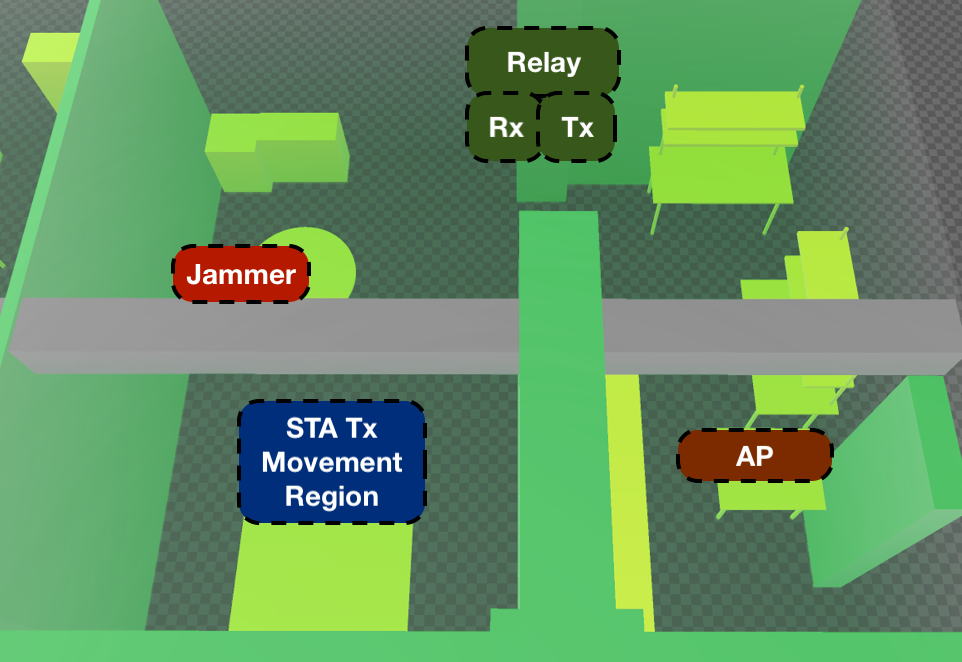}
    \caption{Recreated 3D \ac{dt} of the experimental environment employed for simulations, reproducing all structural and spatial features. The model incorporates walls, doors, windows, furniture, and other relevant elements of the setup. Furthermore, the spatial disposition of the \ac{af} relay, transmitters, and receivers is modeled according to the experimental configuration.}
    \label{fig:3dmap}
\end{figure}

\subsection{Simulation Setup}
\label{subsec:setup}
For the simulations, we use a custom MATLAB-based \ac{dt} simulator, so as it can be properly integrated with a MATLAB ray-tracing simulation engine to model realistic signal propagation within the environment. The tracer simulates up to two reflections, the carrier frequency and antenna configurations are those of Sec.~\ref{sec:sub_experimental_setup}. The nodes are deployed in a reconstructed indoor environment with the same structure, and spatial layout as in the experiments, as illustrated in Fig.~\ref{fig:3dmap}, where all relevant environmental elements are reproduced to ensure high fidelity between simulated and experimental scenarios. In particular, the 3D \ac{dt} includes the structural components of the environment as well as the objects present in the setup, and accurately represents the placement and disposition of the \ac{af} relay, transmitters, and receivers according to the experimental configuration. The simulation parameters are reported in Table~\ref{tab:parameters}.

To generate training and testing samples, the transmitter and interferers are randomly placed within the $x$- and $y$-boundaries of the room, defined by $x_{\min}$, $x_{\max}$, $y_{\min}$, and $y_{\max}$. To maintain consistency with the experimental campaign and avoid unrealistic deployments (e.g., nodes positioned under the furniture), the $z$-coordinate of the nodes is fixed. 
The beamforming codebook in the \ac{dt} simulator is also obtained from the experimental hardware via reverse calibration. Specifically, the beam steering directions at 0\textdegree\: azimuth and 0\textdegree\: elevation, ideally identical in theory, are used as reference calibration values. 
Then, we normalize all other beam steering directions in the codebook with respect to this reference to reproduce beam patterns in the \ac{dt} simulator that accurately reflect the actual hardware behavior, and ensuring consistency with the experimental setup.

\begin{figure}[t!]
    \centering
\begin{tikzpicture}

\definecolor{darkgray176}{RGB}{176,176,176}

\begin{groupplot}[
    label style={font=\fontsize{10pt}{12pt}},
    group style={
        group size=1 by 4,
        vertical sep=1.5cm
    },
    width=8cm,
    height=3cm,
    tick align=outside,
    tick pos=left,
    x grid style={darkgray176},
    y grid style={darkgray176},
    y dir=reverse,
    xmin=0, xmax=21,
    ymin=0, ymax=3,
    xtick={0.5,2.5,4.5,6.5,8.5,10.5,12.5,14.5,16.5,18.5,20.5},
    xticklabels={-54.0,-43.2,-32.4,-21.6,-10.8,0.0,10.8,21.6,32.4,43.2,54.0},
    ytick={0.5,1.5,2.5},
    yticklabels={{18},{0},{-18}},
    yticklabel style={rotate=90}
]

\nextgroupplot[
    title={Real signal (at A1)},
    title style={font=\normalsize},
    ylabel={Elevation [$^\circ$]}
]
\addplot graphics [xmin=0, xmax=21, ymin=3, ymax=0]
{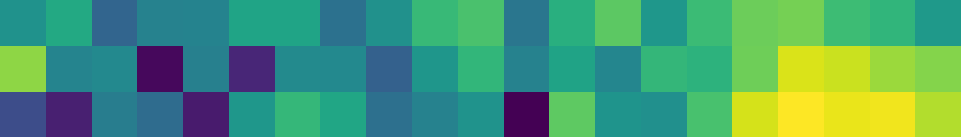};

\nextgroupplot[
    title={Simulated signal (at A1)},
    title style={font=\normalsize},
    ylabel={Elevation [$^\circ$]}
]
\addplot graphics [xmin=0, xmax=21, ymin=3, ymax=0]
{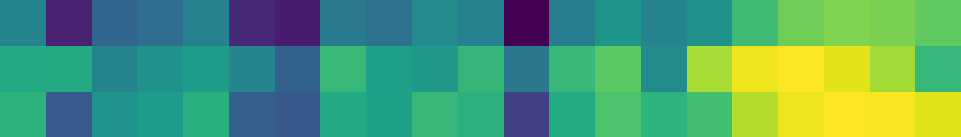};

\nextgroupplot[
    title={Real signal (at B15)},
    title style={font=\normalsize},
    ylabel={Elevation [$^\circ$]}
]
\addplot graphics [xmin=0, xmax=21, ymin=3, ymax=0]
{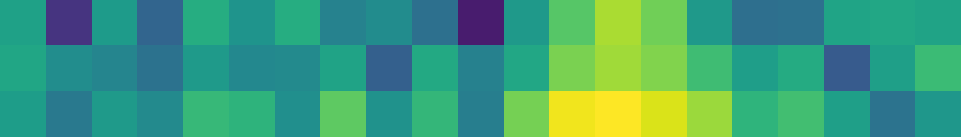};

\nextgroupplot[
    title={Simulated signal (at B15)},
    title style={font=\normalsize},
    xlabel={Azimuth [$^\circ$]},
    ylabel={Elevation [$^\circ$]}
]
\addplot graphics [xmin=0, xmax=21, ymin=3, ymax=0]
{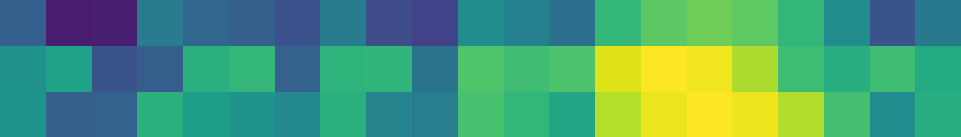};

\end{groupplot}
\begin{axis}[
    hide axis,
    scale only axis,
    xmin=0, xmax=1, ymin=0, ymax=1, 
    at={(group c1r1.north east)}, 
    anchor=north west,
    height=10.5cm, 
    width=0.1,
    colorbar,
    colormap/viridis,
    point meta min=-66.2310890526287,
    point meta max=-20.410979946919,
    colorbar style={
        width=0.2cm, 
        ylabel={Received power (dBm)},
        ylabel style={
            font=\footnotesize,
            at={(1.4,0.5)},
            yshift=-5mm      
        },
        yticklabel style={anchor=west}
    }
]
\end{axis}
\end{tikzpicture}
    \caption{Comparison between simulated and experimental receive power for transmitter positions A1 and B15 (see Fig. \ref{fig:exper_data_pos}). The receive power is shown for each receiving beam. Simulated results are obtained using a ray-tracing propagation model with up to two reflections.}
    \label{fig:comparison_pm0}
\end{figure}

Figure~\ref{fig:comparison_pm0} compares the received signal power at the \ac{ap} measured in the experimental setup with that obtained from the simulated setup across different beam steering directions. A noticeable similarity in the observed patterns is found between the two, supporting the use of a \ac{dt} to emulate realistic wireless scenarios. The performance of the proposed \ac{pao} beamforming framework depends on how accurately the \ac{dt} represents the actual environment, which is supported by this step.
 
\subsection{SL Performance}
\label{sec:sl_perf}
In this section we evaluate the performance of the \ac{sl} step, and measure the \ac{rmse} of the transmitter's and interferers' predicted positions and the \gls{sinr}. The \ac{pao} framework consists of two steps, namely \ac{sl} and \ac{dtao}. Actually, the performance of \ac{dtao} relies on the transmitter's and interferers' positions predicted by the \ac{nn} in the \ac{sl} step. For this reason, we start the evaluation by focusing on the prediction performance of \ac{sl}, measured in terms of the \ac{rmse} of the predicted positions, and the resulting effect on the \ac{sinr}. To train the \ac{nn}, we first generate a synthetic dataset ${\mathcal D}_S$ consisting of 10,000 simulated samples using a \ac{dt} as described in Sec.~\ref{subsec:setup}. Each sample includes the positions of the transmitter and an interferer, along with the corresponding \ac{sinr} values under multiple beam configurations. This dataset is used to pre-train the model. For fine-tuning, we use real-world laboratory data denoted as $\mathcal{D}_T$, which comprises 600 samples collected across 30 distinct locations  as described in Sec.~\ref{sec:sub_experimental_setup}, with 20 repeated measurements per location under consistent environmental conditions.  
During this stage, we freeze the early layers of the \ac{nn}, and only update the deeper, more task-specific layers. Fine-tuning is carried out for a small number of epochs using a reduced learning rate, allowing the model to adapt to real-world conditions while retaining the knowledge acquired from the synthetic data.
We consider $S \in \{3, 5, 7, 11, 21, 63\}$ possible beam directions, to evaluate the trade-off between codebook resolution and accuracy.
Then, the \ac{rmse} is computed as
\begin{equation} \label{mse}
     {\rm RMSE}(\bar{\bm{p}}, \hat{\bm{p}}) =\sqrt{ \frac{1}{||{\mathcal D}_T||} \sum_{ \bar{\bm{p}} \in \mathcal D_T}{ ||\hat{\bm{p}} - \bar{\bm{p}}||^2}},
\end{equation}
where $\bar{\bm{p}}$ and $\hat{\bm{p}}$ are the ground-truth and predicted positions, respectively.
We adopt a leave-one-location-out refinement strategy: at each iteration, data from 29 locations are used for fine-tuning, while the remaining location is used for the testing. This process is repeated for all 30 locations to ensure generalization.

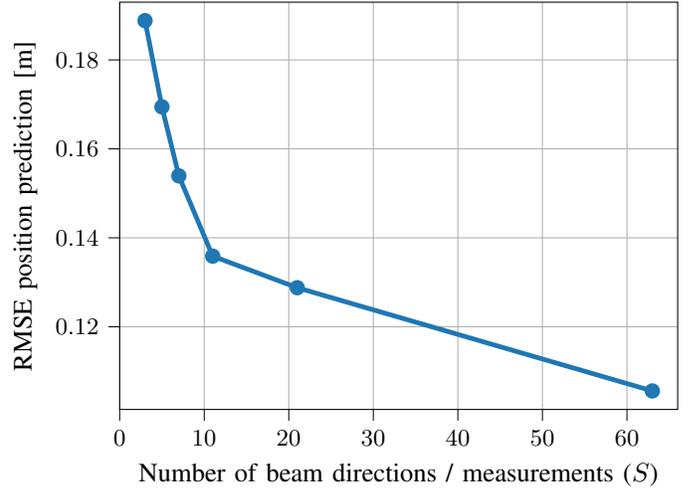
\begin{figure}[t!]
    \centering
\begin{tikzpicture}

\definecolor{color0}{rgb}{0.12156862745098,0.466666666666667,0.705882352941177}

\begin{axis}[
    label style={font=\fontsize{10pt}{12pt}},
    tick label style={font=\fontsize{9pt}{10pt}},
    width=9cm,
    height=7cm,
    tick align=outside,
    tick pos=left,
    x grid style={white!69.0196078431373!black},
    xlabel={Number of beam directions / measurements ($S$)},
    xmajorgrids,
    xmin=0, xmax=66,
    xtick style={color=black},
    y grid style={white!69.0196078431373!black},
    ylabel={RMSE position prediction [m]},
    ymajorgrids,
    ymin=0.101408803, ymax=0.192976917,
    ytick style={color=black}
]
\addplot [draw=color0, fill=color0, mark=*, only marks, line width=1.8pt, mark size=2pt]
table{%
x  y
3 0.18881473
5 0.169441
7 0.15392821
11 0.1358794
21 0.12878036
63 0.10557099
};
\addplot [semithick, color0, line width=1.8pt, mark size=2pt]
table {%
3 0.18881473
5 0.169441
7 0.15392821
11 0.1358794
21 0.12878036
63 0.10557099
};
\end{axis}

\end{tikzpicture}
\caption{RMSE of the predicted positions as a function of the input dimension $S$. Each point corresponds to a independently trained \ac{nn}.}
    \label{fig:rmse_pred}
\end{figure}

Figure~\ref{fig:rmse_pred} shows the \ac{rmse} of the predicted positions as a function of the input codebook dimension $S$, which corresponds to the number of real \ac{sinr} measurements used as input to the \ac{nn} model. 
Each point in the curve represents a separately trained \ac{nn}. 
Notably, beam steering directions are selected by initially fixing the elevation angle to 0\textdegree\ and uniformly scanning the azimuth space. As $S$ increases, additional measurements are taken at higher elevation angles, thereby providing more spatially diverse input features. 
The results show that increasing $S$ significantly improves localization accuracy.
Specifically, the \ac{rmse} decreases steeply up to $S = 11$. Then, for $S>11$, the performance gain becomes marginal, suggesting that using $S^*=11$ beam steering directions provides a good balance between prediction accuracy and the effort required to collect measurements.

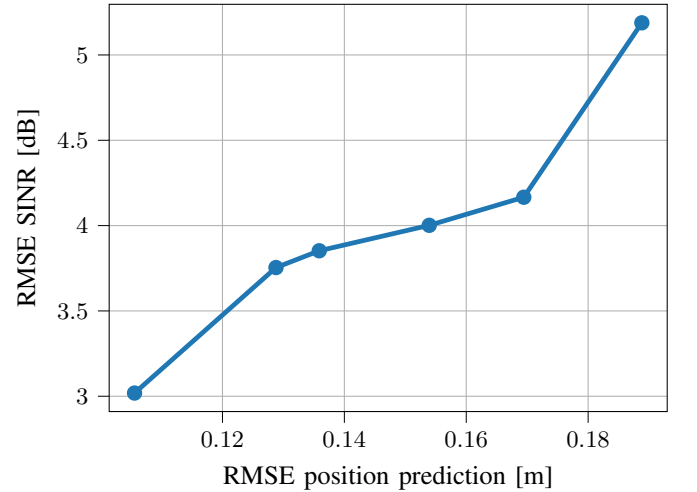
\begin{figure}[t!]
    \centering
\begin{tikzpicture}
\definecolor{color0}{rgb}
{0.12156862745098,0.466666666666667,0.705882352941177}

\begin{axis}[
    label style={font=\fontsize{10pt}{12pt}},
    tick label style={font=\fontsize{9pt}{10pt}},
    width=9cm,
    height=7cm,
    tick align=outside,
    tick pos=left,
    x grid style={white!69.0196078431373!black},
    xlabel={RMSE position prediction [m]},
    xmajorgrids, xmin=0.101408803,
    xmax=0.192976917,
    xtick style={color=black},
    y grid style={white!69.0196078431373!black},
    ylabel={RMSE SINR [dB]},
    ymajorgrids,
    ymin=2.91021977285234, ymax=5.29682555800902,
    ytick style={color=black}
]
\addplot [draw=color0, fill=color0, mark=*, only marks, line width=1.8pt, mark size=2pt]
table{%
x y
0.10557099 3.01870185399582
0.12878036 3.75407653640821
0.1358794 3.85249407371736
0.15392821 4.0016740701586
0.169441 4.16621013372969
0.18881473 5.18834347686554
};
\addplot [semithick, color0, line width=1.8pt, mark size=2pt]
table {%
0.10557099 3.01870185399582
0.12878036 3.75407653640821
0.1358794 3.85249407371736
0.15392821 4.0016740701586
0.169441 4.16621013372969
0.18881473 5.18834347686554
};
\end{axis}
\end{tikzpicture}
\caption{RMSE of the predicted \ac{sinr} (relative to the true \ac{sinr}) vs. the \ac{rmse} of the predicted positions (relative to the true positions).}
\label{fig:rmse_sinr}
\end{figure}

\begin{figure}[t!]
    \centering
    \begin{tikzpicture}
\begin{groupplot}[
    label style={font=\fontsize{10pt}{12pt}},
    group style={
        group size=1 by 3,
        vertical sep=1.5cm
    },
    width=8cm, height=3cm,
    tick align=outside,
    tick pos=left,
    x grid style={white!70!black},
    y grid style={white!70!black},
    y dir=reverse,
    xmin=0, xmax=21,
    ymin=0, ymax=3,
    xtick={0.5,2.5,4.5,6.5,8.5,10.5,12.5,14.5,16.5,18.5,20.5},
    xticklabels={-54.0,-43.2,-32.4,-21.6,-10.8,0.0,10.8,21.6,32.4,43.2,54.0},
    ytick={0.5,1.5,2.5},
    yticklabels={18,0,-18},
    yticklabel style={rotate=90}
]

\nextgroupplot[
    title={\normalsize Simulated signal (at B1)},
    title style={    
        anchor=south,    
        xshift=-2mm      
    },
    ylabel={Elevation [°]}]
\addplot graphics [xmin=0, xmax=21, ymin=3, ymax=0] {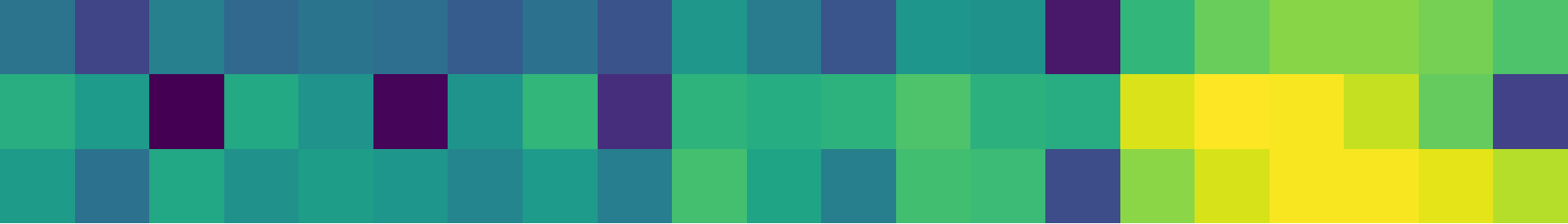};

\nextgroupplot[
    title={\normalsize Signal using predictions of B1 (RMSE$=0.11$m)},
    title style={
        anchor=south,    
        xshift=-2mm      
    },
    ylabel={Elevation [°]}]
\addplot graphics [xmin=0, xmax=21, ymin=3, ymax=0] {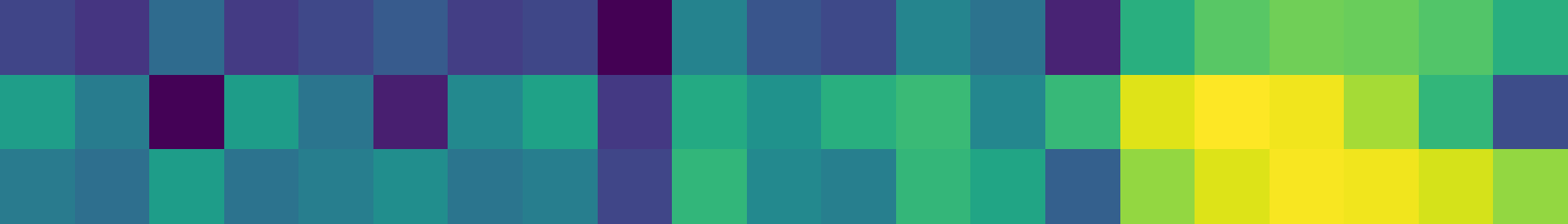};

\nextgroupplot[
    title={\normalsize Signal using predictions of B1 (RMSE$=0.19$m)},
    title style={
        anchor=south,    
        xshift=-2mm      
    },
    xlabel={Azimuth [°]},
    ylabel={Elevation [°]}]
\addplot graphics [xmin=0, xmax=21, ymin=3, ymax=0] {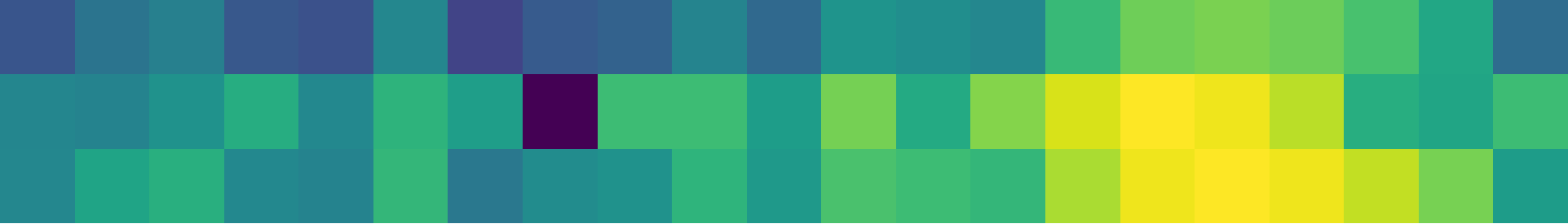};

\end{groupplot}

\begin{axis}[
    hide axis,
    scale only axis,
    xmin=0, xmax=1, ymin=0, ymax=1, 
    at={(group c1r1.north east)}, 
    anchor=north west,
    height=8cm, 
    width=0.1,
    colorbar,
    colormap/viridis,
    point meta min=-68.9,
    point meta max=-20.37,
    colorbar style={
        width=0.2cm, 
        ylabel={Received power (dBm)},
        ylabel style={
            font=\footnotesize,
            at={(1.4,0.5)},
            yshift=-5mm      
        },
        yticklabel style={anchor=west}
    }
]
\end{axis}

\end{tikzpicture}
\vspace{-0.6cm} 
\caption{Received power per beam at position B1. We use the true positions (top) and the predicted positions from experimental data using an \ac{nn}. Prediction errors are expressed in terms of the  \ac{rmse}.}
\label{fig:signal_pos_pred}
\end{figure}

Figure~\ref{fig:rmse_sinr} shows the \ac{rmse} of the predicted \ac{sinr}, computed as the difference between the \ac{sinr} at the predicted positions and the \ac{sinr} at the actual positions, plotted against the \ac{rmse} of the predicted positions. As expected, the \ac{rmse} \ac{sinr} increases with the position error, meaning that improved position prediction accuracy translates into more accurate \ac{sinr} estimation. 
This is consistent with \eqref{sinr_position}, where the \ac{sinr} is modeled as a spatially-dependent function.
Notice that the trend in Fig.~\ref{fig:rmse_sinr} closely resembles that in Fig.~\ref{fig:rmse_pred}.

In addition, Fig.~\ref{fig:signal_pos_pred} reports the receive power per beam at position B1 (see Fig.~\ref{fig:exper_data_pos}). The signal strength is measured either relative to the true positions of the transmitter, or the predicted positions from the NN. The results show that, as the position error (expressed in terms of the \ac{rmse}) increases, the mismatch between the true and \ac{nn}-predicted receive power also increases. Nevertheless, despite the moderate positioning error, the spatial structure of the received power is preserved in both cases, with minor network performance degradation.

\subsection{PAO Convergence Efficiency Demonstration}
\label{sec:interp}

\begin{figure}[t!]
    \centering
    \begin{tikzpicture}
\definecolor{color0}{rgb}{0.12156862745098,0.466666666666667,0.705882352941177}
\definecolor{color1}{rgb}{1,0.647058823529412,0}

\begin{axis}[
    label style={font=\fontsize{10pt}{12pt}},
    tick label style={font=\fontsize{9pt}{10pt}},
  width=9cm,
  height=7cm,
  legend cell align={left},
  legend style={
    font=\fontsize{9pt}{10pt},
    fill opacity=0.9,
    at={(0.97,0.03)},
    anchor=south east,
    draw=white!80!black
  },
  grid=both,
  xlabel={Number of beam directions / measurements ($S$)},
  ylabel={SINR (dB)},
  xmin=0, xmax=66,
  ymin=7.5, ymax=26,
]

\addplot[
  color=color0,
  line width=1.8pt,      
  mark size=2pt,
  mark=*,
  forget plot
]
table {
3 24.0690831823769
5 24.2023040509691
7 24.3023456917476
11 24.493876165651
21 24.4584717930305
63 24.7265279642326
};
\addlegendimage{only marks, mark=*, mark options={fill=color0,draw=color0}}
\addlegendentry{PAO (GA/GBO)}

\addplot[
  color=color1,
  line width=1.8pt,      
  mark size=2pt,
  mark=diamond*,
  forget plot
]
table {
3 8.5319332735066
5 21.4696380117834
7 23.1261552463939
11 23.6713720680028
21 24.0641619023945
63 24.7736186421283
};
\addlegendimage{only marks, mark=diamond*, mark options={fill=color1,draw=color1}}
\addlegendentry{Benchmark}

\end{axis}
\end{tikzpicture}
\caption{SINR of the proposed \ac{pao} framework vs. an interpolation-based benchmark using acquired data, as a function of the input dimension $S$ (i.e., the number of real beam measurements used per sample).}
\label{fig:interpolation}
\end{figure}
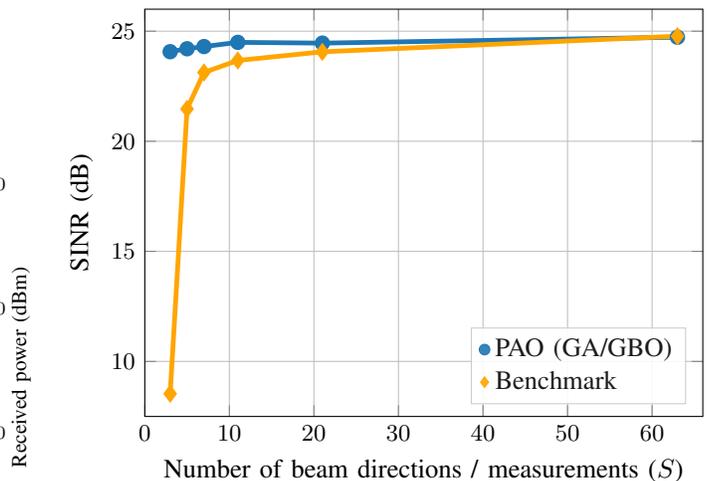

For the simulator, we use a ray-tracing propagation model with up to two reflections. First, in Fig.~\ref{fig:interpolation} we compare the performance of the proposed \ac{pao} framework with a classical linear grid interpolation-based approach, which operates on measurements obtained from a predefined codebook of relay configurations. These measurements start with an initial codebook configuration, followed by incrementally activating neighboring configurations in the grid, each requiring a new measurement by the relay, enabling gradual linear interpolation across the measured \ac{sinr} values to estimate continuous positions, in terms of the \ac{sinr} obtained with the best \ac{af} relay steering direction.
Notice that the interpolation method operates directly on ground-truth positions obtained from the simulations, and therefore does not rely on learning models; so, it serves as a heuristic optimization benchmark for \ac{pao}. In contrast, \ac{pao} operating with experimental data, is source of a prediction error compared to the interpolation approach.

The results show that \ac{pao} converges to a better solution significantly faster, and requires fewer measurements to determine an optimal \ac{af} relay configuration. Eventually, it returns a higher \ac{sinr}. This improvement is attributed to the fact that \ac{pao} uses a \ac{dt} simulator that emulates the wireless environment, enabling beamforming optimization without relying on dense experimental data. In contrast, the interpolation method only depends on ground-truth positions, and requires more data to achieve comparable performance.

\subsection{Optimization Workflow Efficiency}
\label{sec:bl_workflow}

\begin{figure}[t!]
    \centering
    \includegraphics[width=\linewidth]{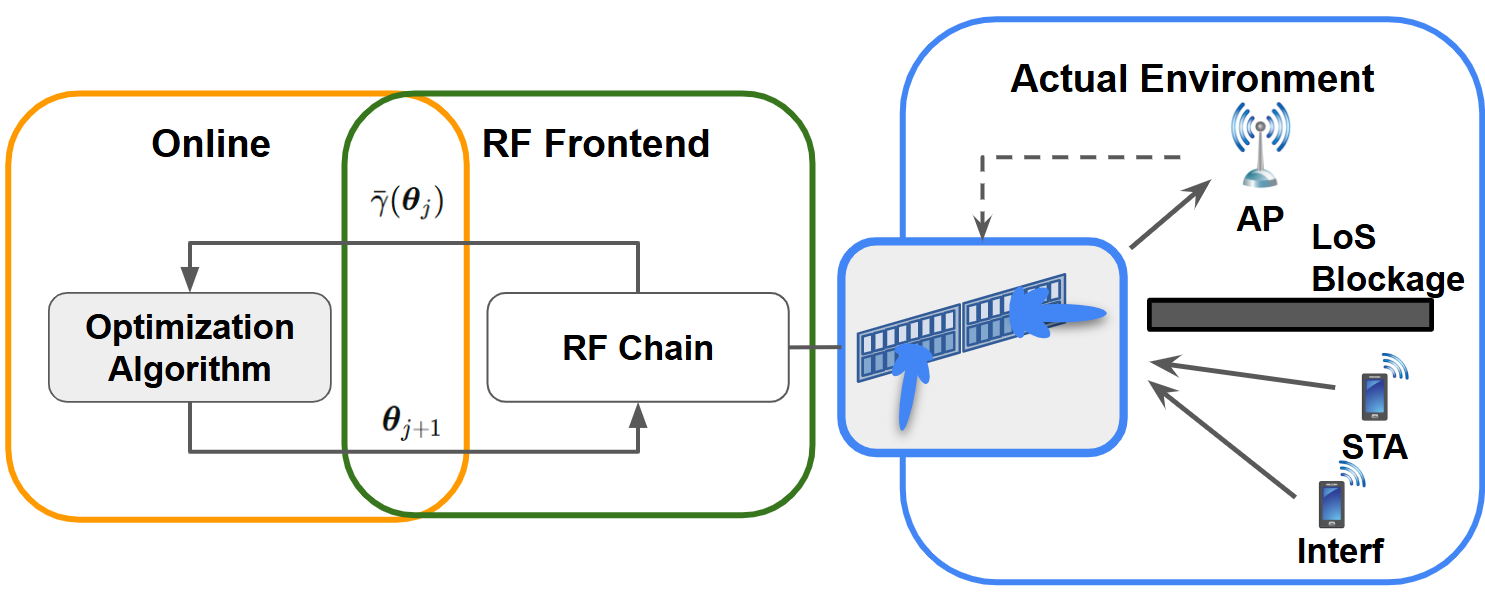}
    \caption{Baseline workflow: direct application of \ac{ga}/\ac{gbo} to the real setup, requiring dense measurements at each optimization step.}
    \label{fig:baseline_workflow}
\end{figure}

Ideally, the baseline would rely entirely on real-world power measurements acquired with the experimental setup described in Sec.~\ref{sec:sub_experimental_setup}. However, the experimental dataset $\mathcal{D}_T$ contains only a limited number of samples. To increase the number of samples, we also consider synthetic data generated by the emulated environment in the \ac{dt}, which has previously demonstrated high fidelity in reproducing real-world behavior (see Fig.~\ref{fig:comparison_pm0}). In this context, the beamforming angles \(\theta_{Az}\) and \(\theta_{El}\) are optimized over a continuous search space \([0, \pi]\), which cannot be directly supported by the hardware implementation based on a quantized codebook.

As shown in Fig.~\ref{fig:baseline_workflow}, the baseline applies \ac{ga} or \ac{gbo} directly to the real setup, which requires extensive measurements from the environment at every optimization step. In contrast, \ac{pao} (see Fig.~\ref{fig:model}) does not apply the optimization algorithms directly on the real hardware. Instead, \ac{pao} first predicts the channel conditions and user positions using the \ac{dt}, and then performs most of the \ac{ga} or \ac{gbo} optimization within this simulated environment. Only a limited number of real-world measurements are required to further refine these predictions.

\begin{figure}
    \centering
    \input{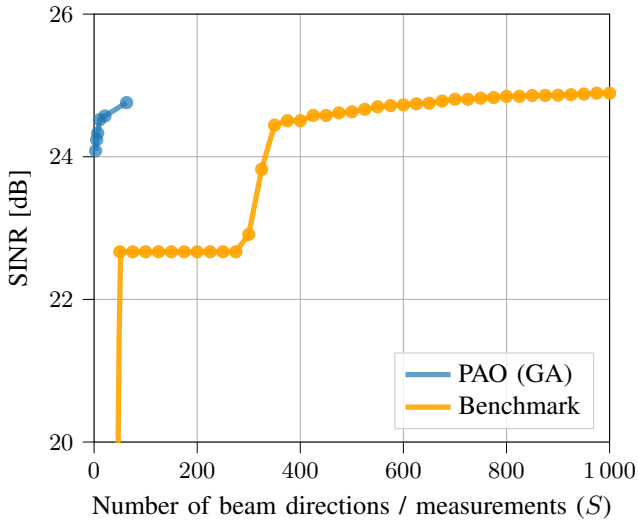}
    \caption{SINR of the proposed \ac{pao} framework with \ac{ga} optimization, vs. a benchmark using only real data, as a function of the input dimension $S$ (i.e.,
the number of real beam measurements per sample).}
    \label{fig:gaopt}
\end{figure}

In Figure~\ref{fig:gaopt} we plot the \ac{sinr} using \ac{pao} with \ac{ga} optimization. 
We see that \ac{pao} achieves relatively high \ac{sinr} values after only a limited number of real-world measurements ($S<63$). This efficiency is due to the fact that most of the optimization is performed within the \ac{dt} simulator, reducing the need for costly measurements. The orange line represents the baseline without \ac{pao}, which typically requires significantly more measurements, and still provides negligible \ac{sinr} improvements. 
We observe an initial sharp increase of the \ac{sinr}, followed by more gradual improvements as additional measurements are considered.
This trend reflects the typical behavior of \ac{ga}-based optimization, where the algorithm first explores a wide range of possible solutions (exploration), and then gradually refines the best candidates (exploitation), resulting in slower but consistent improvements as it converges toward an optimal beamforming configuration.

\begin{figure}
    \centering
    \input{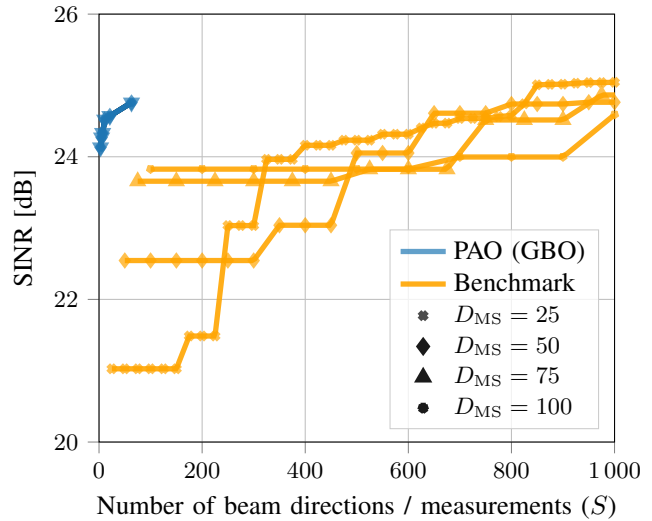}
    \caption{SINR of the proposed \ac{pao} framework with \ac{gbo} optimization as a function of $D_{MS}$, vs. a benchmark using only real data, as a function of the input dimension $S$ (i.e.,
the number of real beam measurements per sample).}
    \label{fig:fmsopt}
\end{figure}

Then, in Fig.~\ref{fig:fmsopt} we consider \ac{pao} with \ac{gbo} optimization, for $D_{MS}\in\{25,50,75,100\}$. Each value of $D_{MS}$ indicates the number of random starting points used in the optimizations. As a benchmark, instead of using the full \ac{pao} framework, we use only the \ac{gbo} algorithm applied directly to the actual environment.
Unlike with \ac{ga}, we now observe multiple, smaller, abrupt jumps in the \ac{sinr}, particularly as $D_{MS}$ increases. This behavior arises because the \gls{gbo} algorithm evaluates several candidate solutions simultaneously and selects the best-performing one at each iteration, resulting in sudden improvements whenever a superior solution is discovered among the parallel runs.
From these results, two distinct optimization behaviors emerge. The \ac{ga} method undergoes an initial exploratory phase with notable \ac{sinr} improvements, followed by a phase of gradual and continuous refinement as it converges toward an optimal beamforming solution. On the other hand, the \ac{gbo} method produces a more irregular progression, characterized by moderate performance jumps corresponding to the selection of the best candidate from multiple parallel optimization paths. Increasing the number of multistarts $D_{MS}$ amplifies this effect by increasing the probability of finding better initial solutions early in the optimization process.

\section{Conclusions}
\label{sec:concs}
In this paper, we proposed a novel \ac{pao} framework for beamforming design in \ac{af} relay-assisted \ac{mimo} systems. Our proposed method operates in two stages. The first stage leverages a \ac{nn} trained to predict the locations of both target and interfering transmitters solely from received power measurements, neglecting the need for explicit \ac{csi}. The second stage employs a \ac{dt}-assisted optimization algorithm that iteratively refines the relay’s beamforming direction using the predicted device positions to maximize the \ac{sinr} at the intended receiver. 
Simulation results demonstrate the effectiveness of the proposed approach in balancing the trade-off between transmitter position prediction accuracy and beamforming performance. Furthermore, integrating the \ac{dt} significantly reduces the need of real-world measurements, while maintaining robust interference mitigation even under localization uncertainties. Therefore, the \ac{pao} framework can be seen as a promising solution for dynamic and interference-rich wireless environments.

\bibliographystyle{IEEEtran}
\bibliography{DT_RIS_bib}

\end{document}